\documentclass{article}

\usepackage{arxiv}

\usepackage[utf8]{inputenc} 
\usepackage[T1]{fontenc}    
\usepackage{hyperref}       
\usepackage{url}            
\usepackage{booktabs}       
\usepackage{amsfonts}       
\usepackage{nicefrac}       
\usepackage{microtype}      
\usepackage{lipsum}		
\usepackage{amssymb} 
\usepackage{amsmath} 
\usepackage{graphicx} 
\title{Possibility of suppression of the formation of solute-enriched clusters}

\author{
  A.~V. Subbotin, M.~I. Gurevich, A.~A. Kovalishin, P.~A. Likhomanova\thanks{CONTACT Email: likhomanovapa@gmail.com} \\
  NRC 'Kurchatov Institute', 123182, Russia, Moscow, Academician Kurchatov Square, 1\\
}

\begin{document}
\maketitle

\begin{abstract}
Based on the earlier proposed mechanism of formation of solute-enriched clusters in irradiated pressure vessel steels, it has been demonstrated that it is possible to suppress this process by the method of light alloying of PVS with a chemical element with specially selected properties such as solubility, activity, and mobility. It is shown that it is possible to change the thermal spike molten zone solidification mechanism and suppress the "solute drag" leading to the formation of solute-enriched clusters. It must be noted that such a solute-enriched cluster formation mechanism is appropriate for alloying elements with low solubilities---it fits for B.C.C. steel-metal of PVS, but does not for the F.C.C. austenitic steels.
\end{abstract}


\section{Introduction}
In the present study, we continue the consideration of one of the two processes causing radiation-induced embrittlement of PWR-type reactor vessel steels, namely grain body hardening\cite{1987JNuM..148..107W}, \cite{1995JNuM..225..196B}, \cite{1996JCCC}, \cite{1997JPV}, \cite{2001JNuM..298..211C}.
Through numerous experimental studies, it was concluded that radiation-induced generation of small-sized ($d \cong 2 \div 4$ nm) quasi-spherical clusters enriched with solute elements $Cu, Mn, Si$, and $Ni$ considerably contributes to radiation-induced grain body hardening \cite{1995JNuM..225..192A}, \cite{1995JNuM..225..225A}, \cite{2000MC}.

Reference \cite{2016JNuM..477..193S} proposed and partly explicated the radiation-induced generation mechanism of solute-enriched cluster ensembles, which involves several stages.

$\alpha$) Thermal Spike Stage


Upon interaction with the lattice, a neutron with sufficient energy generates a primary knocked-on atom (PKA), which in its turn produces a collision cascade and a residual energy dissipation zone. Thermalization results in the generation of a thermal spike in a highly overheated (up to a temperature $T > T_{m}$, where $T_{m}$ is melting temperature) localized zone for $\sim 10^{-11} \div 10^{-12}$ s. The PKA energy interval 

$40 eV \geq E_{PKA} \geq 20 eV$

is limited by the possibility of molten zone generation (for the lower limit) and the subcascading process (for the upper limit), and is interesting on account of the following considerations.

$\beta$) Molten Zone Evolution Stage

A viable liquid phase nucleus can appear as a result of a fluctuation produced in the overheated zone and quickly grow to sizes conditioned by the thermal spike dispersed in the space. The condition for the stoppage of the development of the molten zone  is

$R_{l}^{max}=R(T_{(t)}=T_{m})$

where $R_{l}^{max}$ is the maximum radius of the molten zone; $R(T_{(t)}=T_{m})$ is the thermal spike zone radius, with $T\geq T_{m}$; $t_{0}$ is the duration of development of the molten zone.

The duration of liquid phase generation corresponding to the typical parameters of the thermal spike does not exceed $10^{-12}\div 10^{-13}$ s.

$\gamma$) Solidification and Solute Drag

The thermal spike and the molten zone contained in it are present in the surrounding medium (metal) that has the temperature $\overline{T}$

$\overline{T} \approx \frac{T_{m}}{3}$ $(\overline{T} \approx 580K)$


Natural dispersion of the thermal spike suggests that the thermal spike zone ($T \geq T_{m}$) should begin contracting fast and become smaller than the molten zone. In reality, contraction of the zone corresponding to $T \geq T_{m}$ is mainly determined by the thermal conductivity. The molten zone contraction (sol-liq boundary motion) is governed by the mechanism of atom-by-atom liquid phase connection to the solid phase, where the main parameter is the atomic diffusion coefficient in the liquid phase, which is approximately two orders lower than the thermal conductivity coefficient. It indicates that the sol-liq interface must lag the $R_{t}(T(t_{+})=T_{m})$ interface, and the molten zone attains the purely undercooled condition. 




Even considering the energy released at the sol-liq interface as the melting enthalpy, numerical solution of the corresponding equation of conductivity reveals that the entire molten zone attains the undercooled condition at $T \cong \overline{T}$ for times $\sim n \times 10^{-12}$ s (where $n < 10$) \cite{PC}.

Further evolution of the molten zone in the undercooled condition can result in different scenarios, depending on the liquid phase composition.

In effect, the development of the liquid phase depends on the competition between two processes:

Solidification by way of the solid phase absorbing atoms from the liquid phase---this process is characterized by smooth (i.e. without any discontinuity) movement of the sol-liq interface till the end of the process. In this case, the lifetime of the liquid phase $\tau _{l}$ is determined by the relation


$R_{l}^{max}=\int_0^{\tau_{l}} V(C_{a},C_{b})dt$


where $V(C_{a},C_{b})$ depends on the near-interface concentrations $C_{a}$ of the solute ''$a$'' and $C_{b}$ of the solvent ''$b$'' in the equation for the liquid phase boundary motion rate. Estimations show that (see below)

$\tau_{l} \sim 10^{-8} \div 10^{-9} s$

for the simplest case of the ''fastest'' one-component (solvent ''$b$''-$Fe$) melt.




Solidification through generation of viable (supercritical) nuclei of the solid phase in the liquid phase by fluctuation and further growth of the nuclei is a competing process. 

In this case, movement of the sol-liq interface must become chaotic and the solute drag must stop.

It is worth noting that the ''solute drag'' mechanism is observed owing to the difference in the solubility of the dissolved element ''$a$'' between the liquid and solid phases.

$E_{sol,a}^{s}-E_{sol,a}^{l} > kT$



The driving force of the solute drag mechanism is the solidification process (release of melting enthalpy), which occurs in the presence of stable motion of the sol-liq interface.

Estimations of the case considered in \cite{2016JNuM..477..193S} for $Cu$ solution in $Fe$ solvent at $D_{Cu} \gg D_{Fe}$, where $D_{Cu} and D_{Fe}$ are the diffusion coefficients in the liquid phase, show that the $Cu$ solution exerts a rather small influence on the sol-liq interface up to the late stages.



In this case, the expected time of nucleation of the solid phase exceeds $\tau_{l}$, which indicates that the solidification is due to the interfacial motion. Stable movement of the interface enables the solute drag mechanism and the formation of solute-enriched clusters.

The presence of a spike in the solubility of component ''$a$'' at the sol-liq interface suggests that it is the atoms of ''$b$'' that are mainly absorbed by the solid phase. 




If the atoms ''$a$'' and ''$b$'' have comparable mobilities, absorption priority must lead to an increase in the near-surface concentration of component ''$a$'' (from the liquid phase side) and, consequently, to a decrease in the solidification velocity due to the absorption of the solvent ''$b$''. It is a result of the decrease in the mobility of component ''$b$'' as the concentration of ''$a$'' increases.
 
An increase in $\tau_{l}$ increase the probability of spontaneous generation of a solid phase in the liquid phase, which must destruct the solute drag mechanism and suppress the generation of solute-enriched cluster ensembles.

Finally, the similar probability of sol-liq interface ''poisoning'' is substantiated by the solution of the diffusion equation with regard to the component ''$a$'' (see (3) and (56)) for two different cases. 

$\frac{\partial^{2} F} {\partial C_{a}^2} > 0$ - stable in terms of solution composition for all concentration ranges.

$\frac{\partial^{2} F} {\partial C_{a}^2} < 0$ - unstable in terms of composition in certain concentration intervals (spinodal decomposition).




Here, $F$ is the Helmholtz free energy of the solution.

The boundary conditions and the motion rate of the boundary are obtained from an analysis of the movement of ''$a$'' and ''$b$'' atoms through the sol-liq interface.

Therefore, the liquid phase evolution is described by a self-consistent solution of the diffusion equation (56) for $C_{a}$, with the moving boundary and the boundary conditions depending on $C_{a}$.

It is shown in \cite{2016JNuM..477..193S} that for the case of $Fe-Cu$ the new model of "solute drag" processes explains the available regular patterns of the solute-enriched clusters obtained in the experiments:

- Concentration of cooper atoms in the solute clusters

- Linear dimensions of the solute clusters

- Concentration pattern at the solute cluster boundary

- Residual concentration of the solute atoms outside the solute cluster domain after the solid-liquid interface transit

-Kinetics of accumulation of the solute clusters under neutron irradiation.

The typical time of the development of one solute-enriched clusters is $\tau_{l} \gtrsim 10^{-8}$ s.

It is important to note that the empirically derived and widely used dose dependencies of the embrittlement process for PWR- and WWER-type reactor PVs are based on determening doses of neutrons with energies of either $E_{n} > 1.0 Mev$ (for PWR) or $E_{n} > 0.5 Mev$ (for WWER).

This is evidence that the processes initiated by high-energy neutrons are of significance, which in our opinion, is in favour of the model. 

\section{Boundary Conditions and the Motion Rate of the Boundary}


The following relation is true for both the solid and liquid phases

\begin{equation}
C_{a}^{s,l}+C_{b}^{s,l}+C_{\nu}^{s,l} = 1
\end{equation}


where $C_{a}^{s,l}$ and $C_{b}^{s,l}$ are the solute and solvent concentrations in the solid ($s$) and liquid ($l$) phases.

$C_{\nu}^{s,l} \ll 1$ - concentration of vacancies in the respective phases.


A vacancy in the liquid phase is defined as a local fluctuation in the density (of the order of the atomic volume). Such a definition can be used only for a highly undercooled condition, similar to the case being considered ($T \cong T_{m}/3$) \cite{1959JChPh..31.1164C}, \cite{1988_Ox}, \cite{1997DD}.


Darken's equation (2) is suitable for alloys containing substitutional alloying elements without any assumption of the structural aspects of the material \cite{1948AIME}. According to the presently accepted Darken's equation for the chemical interdiffusion coefficient,

\begin{equation}
\widetilde{D} = (C_{b}D_{a}^{*}+C_{a}D_{b}^{*}) (1+\frac{dln \gamma}{dlnC})
\end{equation}



Here, $D_{a}^{*}$ and $D_{b}^{*}$ are the tracer diffusion coefficients for the components ''$a$'' and ''$b$'', respectively.

The ratio of the activity coefficients in the two-component solution is used in (2): 

$\frac {\partial ln \gamma_{a}} {{\partial ln C{a}}} = \frac{\partial ln \gamma_{b}} {{\partial ln C{b}}}$


where the activities are \cite{1964RPPh...27..161H}, \cite{1972CP}

$a_{a}=\gamma_{a} C_{a}$

$a_{b}=\gamma_{b} C_{b}$


Thus, it is possible to determine the $C_{a}^{l}$ in the liquid phase in the local coordinate system \cite{1972CP},\cite{2004CG} :

\begin{equation}
\nabla (\widetilde{D} \nabla C_{a}^{l}) = \frac{\partial C_{a}^{l}}{\partial t}
\end{equation}


Equation (3), but in the laboratory coordinate system (see (56)), is used for analyzing the quasi-spherical zone of the liquid phase with a moving sol-liq interface. If $T_{l} < T_{m}$, the thermodynamically conditioned process of atomic transition from the liquid to the solid phase is maintained, which decreases the number of atoms in the liquid phase 

$\frac {\partial} {\partial t} (n_{a}+n_{b})<0,$




and, consequently, the sol-liq boundary motion.

The boundary motion rate is determined from the self-consistent solution of equation (56) and the boundary condition which is the function $C_{a}^{l}(R_{(t)})$ in the near-boundary zone (see below).

Let us use the known expression for the phase growth rate for the one-component case \cite{1975AM}, \cite{VCH}, \cite{CRC}, \cite{1958ASM}

\begin{equation}
V = f \frac {6 a_{0} D'}{l^{2}} \left\lbrace 1 - exp \left(- \frac {\partial G}{\partial n} / kT\right) \right\rbrace
\end{equation}


where $D'$ is the diffusion coefficient which considers the jump from the liquid phase to the solid one

$D' = D exp(-\frac{\widetilde S}{k}),$


$D$ is the diffusion coefficient in the liquid phase, $\widetilde{S}$ is the entropy contribution due to the atomic transition from the liquid phase to the solid phase (for metals, $\frac{\widetilde S}{k} \geq 1$ \cite{1958ASM}; $a_{0}$ is the interatomic distance (average for the liquid phase); $l$ is the atomic jump length attributable to the sol-liq boundary transition ($l \sim 2a_{0}$); $f$ - interface roughness - $f \cong 0.25$ for metals (atomic connection; centre concentration); $G(n,N)$ is the Gibbs free energy of the system including the liquid phase and the surrounding solid body:

\begin{equation}
G(n,N)=G^{l}(n)+G^{s}(N)+E^{surf}(n)+E^{el}(n,N)
\end{equation}

where $G^{l}(n) = \mu^{l} n$;

$G^{s}(N) = \mu^{s} N$;

$E^{surf}(n)=s_{0} \overline{\sigma} n^{\frac{2}{3}}$, the interface (surface) energy;

$E^{el}(n,N)$ is the elastic stress energy (see \cite{2016JNuM..477..193S});

$\mu^{l}$ and $\mu^{s}$ are the chemical potentials of the atoms in the liquid and solid phases;

$\overline{\sigma}$ - solid-liquid interfacial energy per atom ($\overline{\sigma} \cong 7.5*10^{-7} eV/at$, for iron);

$s_{0} = 3 (\frac{4 \pi} {3})^{\frac{2}{3}}$;


$n$ and $N$ are the numbers of atoms in the liquid and solid phases, respectively.



It should be noted that the analysis was conducted by taking into account a considerably higher mobility of the atoms in the liquid phase ($\sim 10^{2}$), which enables consideration of the concentrations $C_{a}$ and $C_{b}$ in the solid phase only in the surface layer.

In the two-component case under consideration, the expression for the solid-liquid motion rate $V$, corresponding to the decrease in the number of atoms in the liquid phase, must contain two contributions (see fig. 1).




Thermodynamically conditioned flux of ''$b$'' solvent atoms:

\begin{equation}
I_{b} = f \frac {6 a_{0}}{l^{2}} D'_{b} C_{b}^{l}(R) \left\lbrace exp \left(- \frac {\mu_{b}^{s}-\mu_{b}^{l}}{kT} \right) - 1\right\rbrace
\end{equation}

where $I_{b} > 0$ is the atomic flux entering the solid phase.


Equation (6) is written by neglecting the small surface term $\partial E^{surf}(n) / \partial n$ and considering the fact that $\partial E^{el}(n) / \partial n$ makes sense only at the later stages \cite{2016JNuM..477..193S}:

\begin{equation}
-\frac{\partial G}{\partial n_{b}} \equiv \Delta G_{n_{b} \rightarrow n_{b}-1} \approx  \mu_{b}^{s} - \mu_{b}^{l}
\end{equation}

$\mu_{b}^{l} = kT ln(a_{b}^{l})+\delta E_{b}^{l}$;

$\mu_{b}^{s} = kT ln(a_{b}^{s})+E_{sol,b}^{s}$;

$\delta E_{b}^{l} = \widetilde{h_{b}} \left(\frac{T_{m}-T}{T_{m}}\right)$;

$\widetilde{h_{b}}$ is the melting enthalpy per ''$b$'' atom;

$E_{sol,b}^{l,s}$ are the solubilities of ''$b$'' atoms in the solid and liquid phases of the solvent, and are neglected;

$a_{b,a}^{l}$ and $a_{b,a}^{s}$ are the chemical activities in the liquid and solid phases,  $C_{b}^{l}(R)$  is the near-boundary concentration.


The flux of atoms ''$a$'' has the form

\begin{equation}
I_{a} \approx f \frac {6 a_{0}}{l^{2}} D'_{a} C_{a}^{l} exp \left(- \frac {\mu_{a}^{s}-\mu_{a}^{l}}{kT}\right);
\end{equation}

where as well as for component ''$b$''
 
$-\frac{\partial G}{\partial n_{a}} \approx -\mu_{a}^{l}+\mu_{a}^{s} = ln \frac{a_{a}^{s}}{a_{a}^{l}}+E_{sol,a}^{s}-\widetilde{h_{a}} \left(\frac{T_{m}-T}{T_{m}}\right)$

$\widetilde{h}_{a}$ is the melting enthalpy per atom of ''$a$'' in the solvent phase;

$E_{sol,a}^{s}$ - solubility of ''$a$'' atoms in the solid phase of the solvent - from here on, we neglect the changes in these values with the change in $C_{a,b}^{l}$ and $C_{a,b}^{s}$; $C_{a}^{l}(R)$ is the near-boundary concentration.

Absence of the term describing the transition of ''$a$'' atoms from the solid phase to the liquid phase in equation (8) is explained by two conditions that are simultaneously met:


     For the near-boundary zones of the solid phase, $C_{b}^{s} \gg C_{a}^{s}$(see below);
    
     $\widetilde{D}^{s} / \widetilde{D}^{s} \leq 10^{-2}$.



Thus, it is possible to consider $C_{a}^{s}$ as ''instantaneous'' at each moment, which is a result of the jumps from the liquid phase to the solid phase during the solidification analysis. These considerations make it possible to understand that the reverse flux of ''$a$'' component $\sim (C_{a}^{s})^{2}$ at  $C_{b}^{s} \gg C_{a}^{s}$, and the sol-liq interface is a kind of ''Maxwell demon'' for ''$a$'' component.

Using (6)-(8) we obtain the rate of decrease in the total quantity of atoms in the liquid phase, which is equivalent to the boundary motion rate $V$.

\begin{equation*}
V = I_{a}+I_{b} = f \frac {6 a_{0}}{l^{2}} D_{b} C_{b}^{l} \left(\frac{a_{b}^{l}}{a_{b} ^{s}} \right)  exp \left(\frac {\delta E_{b}-\widetilde{s}T}{kT}\right) \times 
\end{equation*}

\begin{equation}
\times \left\lbrace \left(1 - \frac{a_{b}^{s}}{a_{b}^{l}} e^{-\frac{\delta E_{b}}{kT}}\right) + \frac{D_{a}^{*}C_{a}^{l}}{D_{b}^{*}C_{b}^{l}} \left(\frac {a_{a}^{l} a_{b}^{s}}{a_{a}^{s} a_{b}^{l}}\right) e^{-\frac{\delta E_{a} +\delta E_{b}}{kT}} \right\rbrace
\end{equation}

where $ \delta E_{a} \equiv E_{sol,a}^{s} - \widetilde {h_{a}}\left(\frac{T_{m}-T}{T_{m}}\right)$




The obtained expressions (6), (8), and (9) are exact enough in the case of a ''narrow'' transition zone, which comprises an approximately one atom thick layer.

In the case under consideration, the transition zone is an atomically diffuse interface, in which the transition from the liquid to the solid occurs over several atomic layers \cite{CRC}, \cite{1958ASM}. The ''diffusion'' interface condition ($\delta R \cong (3\div4) a_{0})$) is $\widetilde{h_{a,b}} / kT_{m} \cong 1$, which was just satisfied for the typical metal parameters used by us:  $kT_{m} \cong 0.15 eV$, $\widetilde{h_{a,b}} \approx 0.16-0.14 eV$.

Accordingly, expression (9) is partly qualitative.


It is worth noting that equation (9) is still formal because it involves a two-component system and cannot be used for describing liquid phase solidification. In the process analysed, the concentrations are not arbitrary, and are functionally connected, conditioned by the kinetics of the process.


Such an analysis based on a balance of the fluxes through the boundary is given below. It should be noted that the case of interest for us $D_{a}^{*} \geq D_{b}^{*}$ is different from that considered in \cite{2016JNuM..477..193S} $D_{a}^{*} \gg D_{b}^{*}$ (using dilute solutions technique); local (near-boundary) zones with increased concentration of component ''$a$'' can be obviously produced at the early stages.


Let us use the method reported in proceedings \cite{1964RPPh...27..161H}, \cite{1972CP},\cite{2004CG}, \cite{1958ASM} for description of the matter transport processes in concentrated solid solutions to obtain the required relation $C_{a}^{s}(R)=f(C_{a}^{l}(R))$.




We suppose that the interfacial zone $\mathcal{B}$ exhibits the same peculiarities of the undercooled liquid phase with regard to the diffusion mechanism.

Let us conduct a thermodynamic analysis of the atomic fluxes in the transition layer $\mathcal{B}$ and its surrounding zones, which are considered to contain undercooled liquid phase with a chemical potential jump, for which the ''free volume'' concept is applicable.
 
The transition layer $\mathcal{B}$ displays a drift with respect to the liquid phase in the presence of two fluxes $J_{a}$ and $J_{b}$ (not implicate for $I_{a}$ and $I_{b}$); the drift rate can be obtained from the condition

\begin{equation}
J_{a}+J_{b}+J_{v} = 0,
\end{equation}

from which $J_{v}=-(J_{a}+J_{b})$ is the drift rate.



All the analyses were simplified by neglecting the differences in the solid and liquid phase densities.

Consequently, in $\mathcal{B}$ zone coordinates, the boundary motion rate can be as follows.

\begin{equation}
\overline{V} = V - (J_{a}+J_{b})
\end{equation}

with respect to a random point in $\mathcal{B}$ zone.


The fluxes of the components ''$a$'' and ''$b$'' with respect to the transition layer $\mathcal{B}$ are as follows.

\begin{equation}
{j_{a}=J_{a} - C_{a}^{l}(J_{a}+J_{b}) \atop j_{b}=J_{a} - C_{b}^{l}(J_{a}+J_{b})}
\end{equation}



In turn, it is possible to write expressions for the fluxes $J_{a}$ and $J_{b}$ initiated by the chemical potential gradients $\nabla \mu_{a}^{s,l}$ and $\nabla \mu_{b}^{s,l}$ appearing at the moving boundary by using Onsager relations  \cite{1964RPPh...27..161H},\cite{1972CP},\cite{2004CG} in the following forms

\begin{equation}
J_{a} = - L_{aa} \nabla \mu _{a}^{s-l} - L_{ab} \nabla \mu _{b}^{s-l} \atop J_{b} = - L_{ba} \nabla \mu _{a}^{s-l} - L_{bb} \nabla \mu _{b}^{s-l} 
\end{equation}

$\mu_{a,b}^{s-l} \equiv \mu_{a,b}^{s} - \mu_{a,b}^{l} ; L_{ab} = L_{ba}$. 


Supposing that the whole process is in a weak nonequilibrium state $(D_{a}^{*} / D_{b}^{*} > 1)$, we use Gibbs-Duhem equation

\begin{equation}
C_{a}^{l} \nabla \mu_{a}^{s-l} + C_{b}^{l} \nabla \mu_{b}^{s-l} = 0
\end{equation}


whence it follows that

\begin{equation}
J_{a} = - D_{a}^{*}C_{a}^{l} \nabla \frac{\mu_{a}^{s-l}}{kT}
\end{equation}

\begin{equation}
J_{b} = - D_{b}^{*}C_{b}^{l} \nabla \frac{\mu_{b}^{s-l}}{kT}
\end{equation}

where

\begin{equation*}
D_{a}^{*} = \left( \frac{L_{aa}}{C_{a}} - \frac{L_{ab}}{C_{b}} \right) kT
\end{equation*}

\begin{equation}
D_{b}^{*} = \left (\frac{L_{bb}}{C_{b}} - \frac{L_{ba}}{C_{a}} \right) kT
\end{equation}


and the chemical potential gradients are approximately  

\begin{equation*}
\frac{\nabla \mu_{a}^{s-l}}{kT} \approx \frac{1}{\delta R} ln \left(\frac{a_{a}^{s}}{a_{a}^{l}} exp (\frac{\delta E_{a}}{kT}) \right)
\end{equation*}

\begin{equation}
\frac{\nabla \mu_{b}^{s-l}}{kT} \approx \frac{1}{\delta R} ln \left(\frac{a_{b}^{s}}{a_{b}^{l}} exp (-\frac{\delta E_{b}}{kT}) \right)
\end{equation}


Using (15)-(18) in (12), we obtain the fluxes ''$a$'' and ''$b$'' in the transition layer $\mathcal{B}$  

\begin{equation*}
j_{a} = - \frac{\widetilde{D}^{*}}{\delta R} C_{a}^{l} ln \left(\frac{a_{a}^{s}}{a_{a}^{l}} exp (\frac{\delta E_{a}}{kT}) \right)
\end{equation*}

\begin{equation}
j_{b} = - \frac{\widetilde{D}^{*}}{\delta R} C_{b}^{l} ln \left(\frac{a_{b}^{s}}{a_{b}^{l}} exp (-\frac{\delta E_{b}}{kT}) \right)
\end{equation}

where
$\widetilde{D}^{*} = C_{b}D_{a}^{*} + C_{a} D_{b}^{*}$


By using the relations (10), (11), (14), and (19), it is possible to write the expression for the boundary motion rate for an arbitrary point of the transition layer $\mathcal{B}$

\begin{equation}
\overline{V}(r)= V(R) + \frac{(D_{a}^{*}-D_{b}^{*})}{\delta R} C_{a}^{l} ln \left(\frac{a_{a}^{s}(R)}{a_{a}^{l}(r)} exp(\frac{\delta E_{a}}{kT}) \right)
\end{equation}


where $V$ has the form of (9) and $R+\delta R \leq r \leq R$.


Our goal is to prove that, owing to the present ''difficulty'' of component ''$a$'' crossing zone $\mathcal{B}$ in the near-boundary zone (in the liquid phase), a component ''$a$'' enriched zone must appear, which should result in component ''$b$'' depletion because of condition (1). The change in the relation between $C_{a}^{l}$ and $C_{a}^{l}$ considerably influences the diffusion coefficients $D_{a}$ and $D_{b}$ because they contain a multiplier

\begin{equation*}
1+\frac{dln \gamma}{dln C}, \quad \textrm{(see (21))}
\end{equation*}



which provides the main contribution to the boundary slowdown.

For this reason, the liquid phase lifetime must increase considerably, which may change the solidification mechanism under certain conditions.


Let us obtain the relation $C_{a,b}^{s} = f(C_{a,b}^{l})$. For this, we formulate the flux balance equations. It is possible to write for component ''$a$''

\begin{equation}
\overline{V}C_{a}^{l}+j_{a} = \overline{V} C_{a,s}
\end{equation}




in $\mathcal{B}$ zone coordinates.

Because of (1), we have the analogous relation for ''$b$''.

Let us analyze (21) by using the relations (9), (19), and (20) and considering the relations between activities and concentrations.

\begin{equation*}
a_{a,b} = \gamma _{a,b} (C_{a,b}) C_{a,b}
\end{equation*}



in different concentration ranges $C_{a}^{l},C_{a}^{s}$ and $C_{b}^{l},C_{b}^{s}$. Let us now perform only a quality analysis by using as in \cite{2016JNuM..477..193S} the $a_{a,b}$ behaviour laws for $Fe-Cu$ solid solution (see \cite{1978net..book.....D}) as an example.

All the expressions are written for the inner (liquid) and outer (solid) surrounding zones of the transition layer $\mathcal{B}$, which require the use of approximations (18).


It is possible to divide all the concentration zones $0 \leq C_{a}^{l,s} \leq 1$ into three ranges (see Fig. 2)


\raisebox{.5pt}{\textcircled{\raisebox{-.9pt} {A}}}


\begin{equation}
 {a_{a}^{l} = \gamma _{a}^{l} C_{a}^{l}}; \quad \quad {a_{b}^{l} = C_{b}^{l}}
\end{equation}

\begin{equation*}
{a_{a}^{s} = \gamma _{a}^{s} C_{a}^{s}}; \quad \quad {a_{b}^{s} = C_{b}^{s}}
\end{equation*}

in the range $0 < C_{a}^{l,s} < C_{a}^{A}$;


\raisebox{.5pt}{\textcircled{\raisebox{-.9pt} {B}}}

\begin{equation}
{a_{a}^{l} = \overline{a_{a}^{l}} \cong 0.9}; \quad \quad {a_{b}^{l} = \overline{a_{b}^{l}} \cong 0.8}
\end{equation}

\begin{equation*}
{a_{a}^{s} = \overline{a_{a}^{s}} \leq 1}; \quad \quad {a_{b}^{l} = \overline{a_{b}^{l}} \leq 1}
\end{equation*}

in the range $ C_{a}^{A} < C_{a}^{l,s} < C_{a}^{B} \cong 0.97$;


\raisebox{.5pt}{\textcircled{\raisebox{-.9pt} {C}}}

\begin{equation}
{a_{a}^{l} = C_{a}^{l}}; \quad \quad {a_{b}^{l} = \gamma _{b}^{l} C_{b}^{l}}
\end{equation}

\begin{equation*}
{a_{a}^{s} = C_{a}^{s}}; \quad \quad {a_{b}^{s} = \gamma _{b}^{s} C_{b}^{s}}
\end{equation*}

in the range $ C_{a}^{B} < C_{a}^{l,s} < 1$.


It is worth noting that the activity-concentration relations for many elements in the solid and liquid solutions in different temperature ranges are highly complicated; in particular, they can have several points of discontinuities. It does not change the qualitative sense of the analysis reported below, but can reveal several ''fronts'' in the concentration distribution.


The use of expression \cite{1948AIME}

\begin{equation}
D_{a,b} = \left( 1 + \frac{\partial ln \gamma}{\partial ln C_{a}^{l}} \right) D_{a,b}^{*}
\end{equation}





for the diffusion coefficient is not strict (see for example \cite{2013MG}, \cite{2002MT}), therefore, the diffusion coefficient decreases in certain concentration ranges, however, it does not vanish for obvious reasons.

Thus, the presence of weak concentration dependences of the activities in certain concentration ranges for concentrated solid solutions leads to decreases in both $D_{a,b}$ and $\widetilde{D}$ in these ranges (see (2)) and the appearance of sufficiently expressed concentration ''fronts'' \cite{1972CP}.

Let us analyze (21) in the form

\begin{equation}
C_{a}^{l} - C_{a}^{s} = - \frac{j_{a}}{\overline{V}},
\end{equation}

where
\begin{equation}
j_{a} = - \frac{C_{b}D_{a}^{*}+C_{a}D_{b}^{*}}{\delta R} C_{a}^{l} ln \left(\frac{a_{a}^{s}}{a_{a}^{l}} exp (\frac{\delta E_{a}}{kT}) \right),
\end{equation}

\begin{equation*}
\overline{V} = I_{a} + I_{b} - I_{dr}, \quad \textrm{from} (20)
\end{equation*}

\begin{equation}
I_{b} = \frac{\alpha}{\delta R} D_{b} C_{b}^{l} \left( \frac{a_{b}^{l}}{a_{b}^{s}} \right) e ^{\frac{\delta E_{b}}{kT}} \times \left\lbrace 1 - \frac{a_{b}^{s}}{a_{b}^{l}} e^{-\frac{\delta E_{b}}{kT}} \right\rbrace ,
\end{equation}

\begin{equation}
I_{a} = \frac{\alpha}{\delta R} D_{a} C_{a}^{l} \left( \frac{a_{a}^{l}}{a_{a}^{s}} \right) e ^{-\frac{\delta E_{a}}{kT}},
\end{equation}

\begin{equation}
I_{dr} = -\frac{D_{a}^{*}-D_{b}^{*}}{\delta R} C_{a}^{l} ln \left( \frac{a_{a}^{s}}{a_{a}^{l}} e ^{\frac{\delta E_{a}}{kT}} \right),
\end{equation}

\begin{equation*}
\alpha \equiv f \frac{6a_{0} \delta R}{l^{2}} e ^{\frac{\delta E_{b}-\widetilde{s}T}{kT}}
\end{equation*}




Taking into account that the initial concentrations of the solute elements are rather small ($\overline{C}_{a}^{l} \sim 3*10^{-3} \div 10^{-2}$), the boundary motion process starts from the stage $\raisebox{.5pt}{\textcircled{\raisebox{-.9pt} {A}}} \raisebox{.5pt}{\textcircled{\raisebox{-.9pt} {A}}}:$ $C_{a}^{A} \geq C_{a}^{l} \geq C_{a}^{s}$, where $C_{a}^{A}\cong 5*10^{-2}$.

The relations given by (22) are valid in this case, and their use in equations (26) - (30) enables us to draw the conclusion that in the case $\raisebox{.5pt}{\textcircled{\raisebox{-.9pt} {A}}} \raisebox{.5pt}{\textcircled{\raisebox{-.9pt} {A}}}$, the rate $\overline{V}$ is determined by the dominant term $I_{b}$.

Equation (26) takes the form

\begin{equation}
C_{a}^{l} - C_{a}^{s} = \frac{1}{\alpha} \frac{\widetilde{D}^{*}}{D_{b}} \frac{C_{a}^{l}C_{b}^{s}}{(1-C_{a}^{l})^2} ln \left(\frac{\gamma_{a}^{s}}{\gamma_{a}^{l}} \frac{C_{a}^{s}}{C_{a}^{l}} e ^{\frac{\delta E_{a}}{kT}} \right)
\end{equation}


Assuming that the process of ''solute drag'' takes place and, consequently, $C_{a}^{l} \gg C_{a}^{s}$, we obtain from (31)

\begin{equation}
C_{a}^{s} \approx \frac{\gamma_{a}^{l}}{\gamma_{a}^{s}} C_{a}^{l} exp \left( - \frac{\delta E_{a}}{kT} + \alpha \left(\frac{D_{b}}{\widetilde{D}^{*}} \right) (1-C_{a}^{l})^{2} \right)
\end{equation}


From this, we obtain by considering that in the case $\raisebox{.5pt}{\textcircled{\raisebox{-.9pt} {A}}}\raisebox{.5pt}{\textcircled{\raisebox{-.9pt} {A}}}$ 

$C_{a}^{l} \ll 1$ , $D_{b} = D^{*}$, $\widetilde{D}^{*} = D_{a}^{*}$,

\begin{equation}
C_{a}^{s} \approx \frac{\gamma_{a}^{l}}{\gamma_{a}^{s}} C_{a}^{l} exp \left( - \frac{\delta E_{a}}{kT} + \alpha \left(\frac{D_{b}^{*}}{D_{a}^{*}} \right) \right)
\end{equation}

                       

Relation (33) determines the connection between $C_{a}^{s}$ and $C_{a}^{s}$ in the range $\raisebox{.5pt}{\textcircled{\raisebox{-.9pt} {A}}} \raisebox{.5pt}{\textcircled{\raisebox{-.9pt} {A}}}$: $C_{a}^{l} \leq C_{a}^{A}$; $C_{a}^{s} \leq C_{a}^{A}$
  
Let us estimate the degree of solute drag for the case $\raisebox{.5pt}{\textcircled{\raisebox{-.9pt} {A}}}\raisebox{.5pt}{\textcircled{\raisebox{-.9pt} {A}}}$.

\begin{equation}
\kappa \equiv \frac{C_{a}^{s}}{C_{a}^{l}} = \frac{\gamma_{a}^{l}}{\gamma_{a}^{s}}
exp \left(- \frac{\delta E_{a}}{kT} + \alpha \left(\frac{D_{b}^{*}}{D_{a}^{*}} \right) \right)
\end{equation}




Therefore, a considerable solute drag process takes place.

Let us now estimate the equilibrium deviation measure.

\begin{equation}
\delta_{eq} = \frac{a_{a}^{s}}{a_{a,0}^{s}}
\end{equation}


where the equilibrium value of the activity in the solid phase at the specified concentration in the liquid phase $C_{a,l}$ is determined from the condition

\begin{equation*}
\Delta \mu_{a}^{s-l} = 0
\end{equation*}

whence

\begin{equation}
a_{a,0}^{s} = \gamma_{a}^{l} C_{a}^{l} exp \left(- \frac{\delta E_{a}}{kT} \right)
\end{equation}


Using (36), we obtain

\begin{equation}
\delta_{eq} = exp \left(\alpha \frac{D_{b}^{*}}{D_{a}^{*}} \right)
\end{equation}




Expression (33) shows obviously that if the boundary sol-liq moves, at least at the stage $\raisebox{.5pt}{\textcircled{\raisebox{-.9pt} {A}}} \raisebox{.5pt}{\textcircled{\raisebox{-.9pt} {A}}}$ $C_{a}^{l}$ must increase; $C_{a}^{s} \ll C_{a}^{l}$ and remains in the range $\raisebox{.5pt}{\textcircled{\raisebox{-.9pt} {A}}} \raisebox{.5pt}{\textcircled{\raisebox{-.9pt} {A}}}$.

The next step should be the consideration of the case $\raisebox{.5pt}{\textcircled{\raisebox{-.9pt} {B}}} \raisebox{.5pt}{\textcircled{\raisebox{-.9pt} {A}}}$ - $C_{a}^{l}$  in the range $C_{a}^{l} > C_{a}^{A}$; meanwhile, $C_{a}^{s}$ remains in the range $C_{a}^{s} < C_{a}^{A}$ ($\raisebox{.5pt}{\textcircled{\raisebox{-.9pt} {A}}}$).



However, before analyzing the case $\raisebox{.5pt}{\textcircled{\raisebox{-.9pt} {B}}} \raisebox{.5pt}{\textcircled{\raisebox{-.9pt} {A}}}$, let us determine the concentration range where the flux of component ''$b$'' - $I_{b}$ dominates (the case where $C_{a}^{l} > C_{a}^{A}$).
 
Let us consider the condition of the fluxes $I_{b}$ and $I_{a}$ being comparable and write them in the form

\begin{equation}
\overline{V} C_{a}^{s} = \overline{V} C_{b}^{s}
\end{equation}


Considering that $C_{a}^{s} \cong C_{b}^{s}$, it is obvious that $a_{a}^{s} = \overline{a}_{a}^{s}$ , $a_{b}^{s} = \overline{a}_{b}^{s}$  (see (23)). On the other hand, it follows from the equality of $I_{b}$ and $I_{a}$ that

\begin{equation}
D'_{a}C_{a}^{l} \left( \frac{a_{a}^{l}}{\overline{a}_{a}^{s}} \right) exp \left(- \frac{\delta E_{a}}{kT} \right) = D'_{b} (1 - C_{a}^{l}) \left( \frac{a_{b}^{l}}{\overline{a}_{b}^{s}} \right)  exp \left(\frac{\delta E_{b}}{kT} \right) 
\end{equation}

 

where we do not yet define $a_{a}^{l}$ and $a_{b}^{l}$ concretely for the $C_{a}^{l}$ and $C_{b}^{l}$ location zone.

It follows from (39) that 

\begin{equation*}
\frac{D'_{a}}{D'_{b}} \frac{\overline{a}_{b}^{s}}{\overline{a}_{a}^{s}} exp \left(- \frac{\delta E_{a}+\delta E_{b}}{kT} \right) = \frac{(1-C_{a}^{l})}{C_{a}^{l}} \frac{a_{b}^{l}}{a_{a}^{l}} 
\end{equation*}

whence

\begin{equation}
\frac{(1-C_{a}^{l})}{C_{a}^{l}} \frac{a_{b}^{l}}{a_{a}^{l}} \ll 1
\end{equation}


Condition (40) implies that

\begin{equation}
   a_{b}^{l} = \gamma _{b}^{l} C_{b}^{l} \atop a_{a}^{l} = C_{a}^{l}
\end{equation}



It suggests that the concentration value $C_{a}^{l}$ at which the fluxes are compared is in zone $\mathcal{C}$ (see (24)). 

Using (40) and (41), we obtain the value $\widetilde{C}_{a}^{l}$ at which the fluxes ''$a$'' and ''$b$'' are compared.

\begin{equation}
\widetilde{C}_{a}^{l} \approx 1 - \left( \frac{D'_{a}}{D'_{b}} \frac{\overline{a}_{b}^{s}}{\overline{a}_{a}^{s}} \frac{1}{\gamma _{b}^{l}} \right) ^{\frac{1}{2}} exp \left(- \frac{\delta E_{a}+\delta E_{b}}{kT} \right)
\end{equation}


For earlier values of the parameters (Appendix 3), 

\begin{equation}
\widetilde{C}_{a}^{l} \approx 1 - 0.25 \times 10^{-2}
\end{equation}


Let us also consider the condition of vanishing flux $I_{b}$, from which the condition for ${C}_{b}^{l}$ follows. Using (28), we obtain

\begin{equation*}
a_{b}^{l} = a_{b}^{s} exp \left( -\frac{\delta E_{b}}{kT} \right)
\end{equation*}


?onsidering that in the developed solute drag process $a_{b}^{s} = C_{b}^{s} \leq 1$, we obtain the value $C_{a}^{*l}$ when $I_{b}$ vanishes:

\begin{equation}
C_{a}^{*l} \cong 1 - exp \left( -\frac{\delta E_{b}}{kT} \right)
\end{equation}


Comparison of (43) and (44) demonstrates that the condition $I_{b} =0$ precedes $I_{b} = I_{a}$, consequently, we shall further consider the case $\raisebox{.5pt}{\textcircled{\raisebox{-.9pt} {B}}} \raisebox{.5pt}{\textcircled{\raisebox{-.9pt} {A}}}$ for the range

\begin{equation*}
C_{a}^{*l} > C_{a}^{l} > C_{a}^{A}
\end{equation*}


By taking into account the fact that $I_{b}$ prevails in the range under consideration and that the activities for $C_{a}^{l}, C_{b}^{l}$ obey (23), and, those for $C_{a}^{s}, C_{b}^{s}$, expression (22), we obtain an expression similar to that of the case $\raisebox{.5pt}{\textcircled{\raisebox{-.9pt} {A}}} \raisebox{.5pt}{\textcircled{\raisebox{-.9pt} {A}}}$: 

\begin{equation}
{C}_{a}^{s} \approx \frac{\overline{a}_{a}^{l}}{\gamma _{a}^{s}}
exp \left(- \frac{\delta E_{a}}{kT} + \alpha \frac{D_{b}}{\widetilde{D}^{*}} \overline{a}_{b}^{l} (1 - C_{a}^{l}) \right)
\end{equation}



It is not difficult to see that in this case $\kappa \ll 1$ i.e. the solute drag case, $C_{a}^{s} \ll C_{a}^{l}$ is realized.

The equilibrium deviation measure in this case has the form 

\begin{equation*}
\delta_{eq} \cong exp \left\lbrace \alpha \frac{D_{b}}{\widetilde{D}^{*}} \overline{a}_{b}^{l} (1 - C_{a}^{l}) \right\rbrace \approx 1 + \frac{D_{b}}{\widetilde{D}^{*}} \overline{a}_{b}^{l} (1 - C_{a}^{l})
\end{equation*}

at $C_{a}^{l} < C_{a}^{*l}$.



It should be noted that a considerable decrease in $D_{b}$ is observed in this zone.

Let us estimate the contribution of the drift term $I_{dr}$ at the stage $\raisebox{.5pt}{\textcircled{\raisebox{-.9pt} {B}}} \raisebox{.5pt}{\textcircled{\raisebox{-.9pt} {A}}}$ in the range $C_{a}^{*l} > C_{a}^{l}$ (see (26-29)). By using (45) for $C_{a}^{s}$, it is easy to demonstrate that

\begin{equation*}
\frac {I_{dr}}{I_{b}} = \frac{D_{a}^{*} - D_{b}^{*}}{\widetilde{D}^{*}} C_{a}^{l} < 1
\end{equation*}


up to the value

\begin{equation}
C_{a}^{**l} \cong \frac{\frac{D_{a}^{*}}{{D_{b}^{*}}}}{2[\frac{D_{a}^{*}}{D_{b}^{*}} - 1]} 
\end{equation}


at which $I_{dr}$ begins prevailing in the expression for $\overline{V}$.



Further, we shall consider the following simple case without the loss of generality of the results:

$\frac{D_{a}^{*}}{{D_{b}^{*}}} \geq 2.4$ i.e. $C_{a}^{**l} \leq C_{a}^{*l}$



Under this condition, expression (45) is valid for $C_{a}^{**l} \geq C_{a}^{l} \geq C_{a}^{A}$.

Considering that the dominant term is $I_{dr}$ in $\overline{V}$ at $C_{a}^{l} > C_{a}^{**l}$, it is possible to write the flux balance equation (26) in this range of concentrations in the form 

\begin{equation}
C_{a}^{l} - C_{a}^{s} = - \frac{j_{a}}{I_{dr}}
\end{equation}


whence it follows that

\begin{equation}
C_{a}^{l} - C_{a}^{s} = \frac{D_{a}^{*} - D_{b}^{*}}{\widetilde{D}^{*}}
\end{equation}


or, in a more convenient form 

\begin{equation}
C_{a}^{s} = \frac{\Omega - 1}{\Omega} C_{a}^{l}
\end{equation}

where $\Omega \equiv \frac{(\beta-1)C_{a}^{l}}{\beta-(\beta-1)C_{a}^{l}}$ , $\beta \equiv \frac{D_{a}^{*}}{D_{b}^{*}}$

Table 5





It becomes obvious from (48), (49), and Table 5 that $C_{a}^{s}$ increases up to values comparable to $C_{a}^{l}$ and, consequently, the solute drag of component ''$a$'' stops at $C_{a}^{**l} > C_{a}^{l}$.

The solute drag measure is $\kappa \cong \frac{\Omega-1}{\Omega}$ in this case, which corresponds to the actual stopping of the process.

The equilibrium deviation measure

$\delta_{eq} \cong \frac{\overline{a}_{a}^{s}}{\gamma_{a}^{l}} exp \lbrace - \frac{\delta E_{a}}{kT} \rbrace \simeq\gg 1$ refers to the short stage of essential deviation from equilibrium.




The obtained expressions for the relation between $C_{a}^{l}$ and $C_{a}^{s}$ and analysis of the equitable solution areas make it possible to write the expressions for the sol-liq boundary motion rate and for the flux of the solute ''$a$'' atoms through the boundary. Taking into consideration the fact that diffusion processes in the solid phase are actually blocked (''frozen''), in comparison to the liquid phase, $\overline{V}$ corresponds to the sol-liq boundary motion rate with respect to the solid phase. 

I. Case where $C_{a}^{l} \leq C_{a}^{A}$ (solute drag):
 
- the sol-liq boundary motion rate with respect to the solid phase is given by

\begin{equation}
\overline{V} \cong \frac{\alpha}{\delta R} D_{b}^{*} C_{b}^{l}
\end{equation}

where $C_{b}^{l} \leq 1$


- the flux of component ''$a$'' from the liquid phase crossing the boundary is

\begin{equation}
I_{a} =  \frac{\alpha}{\delta R} \frac{\gamma_{a}^{l}}{\gamma_{a}^{s}} D_{b}^{*} C_{a}^{l} exp \left\lbrace - \frac{\delta E_{a}}{kT} + \alpha \frac{D_{b}^{*}}{D_{a}^{*}} \right\rbrace
\end{equation}



II. Case where $C_{a}^{**l} \geq C_{a}^{l} \geq C_{a}^{A}$ (solute drag):
 
- the boundary motion rate is

\begin{equation}
\overline{V} \cong \frac{\alpha}{\delta R} D_{b} C_{b}^{l}
\end{equation}

where $1 - C_{a}^{A} \geq C_{b}^{l} \geq 1 - C_{a}^{**l}$


-the flux of component ''$a$'' moving to the solid phase is

\begin{equation}
I_{a} =  \frac{\alpha}{\delta R} \frac{\overline{a}_{b}^{l} \overline{a}_{a}^{l}}{\gamma _{a}^{s}} D_{b} (1-C_{a}^{l}) exp \left\lbrace - \frac{\delta E_{a}}{kT} + \alpha \frac{D_{b}}{\widetilde{D}^{*}} \overline{a}_{a}^{l} (1-C_{a}^{l}) \right\rbrace
\end{equation}



III. Case where $C_{a}^{l} > C_{a}^{**l}$ (solute drag stopped): $C_{a}^{l} \sim C_{a}^{s}$

- the boundary rate is

\begin{equation}
\overline{V} \cong \frac{D_{a}^{*}-D_{b}^{*}}{\delta R} C_{a}^{l} \left(\frac{\delta E_{a}}{kT}-\frac{1}{\Omega} \right)
\end{equation}


-the flux of component ''$a$'' moving to the solid phase is

\begin{equation}
I_{a} \cong \frac{D_{a}^{*}-D_{b}^{*}}{\delta R} \frac{\delta E_{a}}{kT} \left(1 -\frac{1}{\Omega} \right) (C_{a}^{l})^{2}
\end{equation}

\section{Evolution of the Concentration of Solute ''a'' in the Liquid Phase During Solidification}


In the previous section, for obtaining the boundary conditions, we used a local coordinate system connected with an arbitrary element of the volume under drift which was different in each element at the rate

\begin{equation*}
V_{dr} = (D_{a}-D_{b}) \nabla C_{a}^{l} = (D_{a}^{*}-D_{b}^{*}) C_{a}^{l} \nabla \frac{\mu_{a}^{l}}{kT},
\end{equation*}



which is what equation (3) corresponds to.

However, it is necessary to use the laboratory coordinate system for the entire volume of the liquid phase for obtaining the solution for $C_{a}^{l}(r,t)$. It is easy to demonstrate that the transition from (3) to the laboratory system yields a solution for the spherically symmetrical case with a moving boundary:

\begin{equation}
\frac{\partial}{\partial t} C_{a}^{l} = \nabla(D_{a} \nabla C_{a}^{l})
\end{equation}

where $D_{a} = D_{a}^{*} \lbrace 1 + \frac{dln \gamma}{dln C} \rbrace$,

$r \leq R(t)$ is the boundary coordinate.


In the range $r \leq R(t)$, the initial conditions for (56) have the form

\begin{equation*}
 R(t)|_{t=0} = R_{l}^{max} 
\end{equation*}

\begin{equation*}
 C_{a}^{l}(r,t)|_{t=0} = \overline{C}_{a}^{l},
\end{equation*}


and the boundary conditions 

\begin{equation}
\frac{dR}{dt} = - \overline{V} = -\frac{\alpha}{\delta R} D_{b} C_{b}^{l}
\end{equation}


for the boundary motion rate;

\begin{equation}
 - D_{a} \nabla C_{a}^{l} = J_{a}
\end{equation}


Where the flux through the boundary is given by expression (51) for $C_{a}^{l} \leq C_{A}^{l}$ and (53) for $C_{A}^{l} < C_{a}^{l} < C_{a}^{**l}$.


See estimations of the upper limit of $C_{a}^{**l}$ in Table 4.



It is worth noting that, as shown in both (51) and (53), the fluxes through the boundary are negligible in comparison with the flow of the solvent ''$b$'' atoms. Besides, as seen in the obtained solutions, the condition $C_{a}^{l} \cong C_{a}^{**l}$ is not practically reached in physically interesting cases - all the solutions are practically in the left part of the diagram in Fig. 2.

The formulated equation, in combination with the boundary conditions, is a self-consistent non-linear problem, for which an additional external condition is the dependence

\begin{equation}
1 + \frac{dln \gamma}{dln C} \equiv g = \frac{dln a_{l}}{dln C}
\end{equation}


Noting that we do not consider the concrete solute ''$a$'', but rather we set the inverse problem, to demonstrate that it is possible to suppress the solute drag process at certain values of $a(c_{a})$ and $g(c_{a})$,  we introduce approximations rather arbitrarily (Fig. 3): 

\begin{equation}
a(C_{a}^{l}) = \frac{\gamma _{a} C_{a}^{l}}{\left[1 + \left( \frac{C_{a}^{l}}{C_{A}^{l}} \right) ^{k} \right]^{1/k}}
\end{equation}

\begin{equation*}
g(C_{a}^{l}) = \frac{1}{1 + \left( \frac{C_{a}^{l}}{C_{A}^{l}} \right) ^{k} }
\end{equation*}


which correspond to the left part of the diagram for $a(c_{a}^{l})$ (see Fig. 2)

In this case, $ {\bf C_{a} = 0.007, b = 0, Q = 1, K = 150}$.



Our aim is to obtain solutions for rather light alloying which are in line with the ideology of the vessel steels considered (see Appendix 1). 

Let us perform calculations for the case

$D_{a}^{*}/D_{b}^{*} = 3, \overline{C}_{a}^{l} = 1.0 \times 10^{-3}, C_{A}^{l} = 7 \times 10^{-3}, k=150$.



Fig. 4-7 show the evolution of the concentration $C_{a}^{l}$ and the parameters that depend on it. We can observe that the mobilities of both the components ''$b$'' and ''$a$'' start changing considerably when they approach the value $C_{a}^{l} \approx C_{A}^{l}$ (see (25)). It explains the sharp drop in the interfacial boundary motion rate and the increase in the liquid phase lifetime.
 

\section{Solute Drag Interruption}




It has been demonstrated above that stable solute drag appears in the case of liquid phase solidification by accession of atoms from the liquid phase to the interface boundary, which eventually results in the formation of solute-enriched clusters. The case $D_{a}^{*}/D_{b}^{*} \gg 1$ \cite{2016JNuM..477..193S} was considered.

However, it is not the only way of solidification if the parameters of the solute atoms vary. If it is feasible for fluctuations to generate viable nuclei of the solid phase which grow and coalesce in the liquid phase, the process of formation of uniform solute-enriched clusters is interrupted (see Fig. 8).

Here, the formation of a compact zone with solute concentration $C_{a} \sim 30 \div 80 \%$, which can be the centre of pinning, is not feasible anymore.




Let us consider the value of $Z$ which corresponds to the number of viable nuclei that appear in the liquid phase during its lifetime as a result of fluctuations to estimate the feasibility of the solidification mechanism changing.

In the case of $Z < 1$, the mechanism change process is less probable; stable solute-enriched clusters are formed.

In the case of $Z > 1$, the mechanism can be changed and the formation of solute-enriched clusters can be stopped.

\begin{equation}
Z = \int_0^{\tau} \mathcal{J}_{n}(T) \mathcal{V} (t) dt
\end{equation}

where $\mathcal{J}_{n}(t)$ refers to the number of viable nuclei of the solid phase appearing in unit volume per unit time,


\begin{equation}
\mathcal{V} (t) = \frac{4}{3} \pi (R_{max} - \overline{V}t)^{3}
\end{equation}



- the liquid phase volume at time $t$,

$\tau$ is the time for which the present analysis makes sense.

The flux of solid phase viable nuclei ($n > n^{*}$) generated in 1cm$^{3}$, containing $N$ atoms, has the form (see \cite{1942JETP}, \cite{1966AP}, \cite{PhysKin})

\begin{equation}
\mathcal{J}_{n}(t) = 2 \sqrt{-\frac{1}{2 \pi kT} \frac{\partial ^{2}}{\partial n^{2}} \Delta G(n)\vert _{n^{*}}} \times D(n^{*}) \times N \times exp \left\lbrace - \frac{\Delta G(n^{*})}{kT} \right\rbrace ,
\end{equation}


where $n$ is the number of atoms in the nucleus. 

\begin{equation}
n^{*} = 8 \frac{4 \pi}{3} \left(\frac{\widetilde{\sigma}}{\widetilde{h} \theta} \right)^{3}
\end{equation}


- the number of atoms in the critical nucleus (for the chosen parameters $n^{*} = 13$ atoms).

$D(n^{*}) = 16 \pi \frac{\overline{V}}{\Omega^{1/3}} \left(\frac{\widetilde{\sigma}}{\widetilde{h} \theta} \right)^{2}$


- the diffusion coefficient in $n$ - number space in the neighbourhood $n^{*}$

\begin{equation}
\Delta G(n,T) = - \widetilde{h} \left( \frac{T_{m}-T}{T_{m}} \right) n +s_{0} \widetilde{ \sigma} n^{2/3}
\end{equation}




- thermodynamic potential of the solid phase nucleus containing $n$ atoms, $\theta \equiv \frac{T_{m}-T}{T_{m}}$

Taking into account the fact that it is interesting to consider a solute component with a rather high solubility energy in the solid phase and the concentrations $C_{a}$ being rather small, expressions (62)-(64) are written as approximations of the pure solvent. 

Finally, the expression for $\mathcal{J}_{n}(t)$ takes the form

\begin{equation}
\mathcal{J}_{n}(t) = 2 \frac{\overline{V}}{\Omega^{1/3}} \sqrt{\frac{\widetilde{\sigma}}{kT}} \times N \times exp  \left\lbrace - \frac{16 \pi}{3} \left(\frac{\widetilde{\sigma}}{\widetilde{h} \theta} \right)^{2} \frac{\widetilde{\sigma}}{kT} \right\rbrace
\end{equation}


For the selected parameters, we obtain

$\mathcal{J}_{n}(t) \cong 0.7 \times 10^{25} \times \overline{V} \equiv I_{0} \overline{V}$.



Let us analyze by considering concrete examples.

1. Let $\frac{D_{a}^{*}}{D_{b}^{*}} \gg 1$. In this case, the concentration $C_{a}^{l}$ is approximately uniformly distributed in the volume and the migration rates at the boundary and in the volume are equal for both $C_{a}^{l} < C_{A}^{l}$ and $C_{a}^{l} > C_{A}^{l}$. It is possible to write the expression for $Z$ by using relations (66) and (62) with equally changing $\overline{V}$:

\begin{equation*}
Z=I_{0}V_{max}  \frac{R_{max}}{4} \approx 0.45
\end{equation*}


The obtained estimate suggests that the mechanism does not change at equal mobilities both in the boundary neighbourhood and in the volume (i.e. $C_{a}^{l}$ reaches the value of $C_{A}^{l}$ simultaneously in all the volume) - it is the case of stable solute-enriched cluster formation.




2. The case where $\frac{D_{a}^{*}}{D_{b}^{*}} \cong 3; C_{a}^{l} = 7 \times 10^{-3}; \overline{C}_{a}^{l} = 1.0 \times 10^{-3}$ in the approximation for $g(C_{a}^{l})$ (equation (60)), and $k=150$ is used. The solution of equation (56) is presented in Fig. 4.
 
It is possible to draw the following conclusions from the obtained calculation results.

I. The stage at which the concentration ${C}_{A}^{l}$ (curve ($\alpha$)) is reached at the boundary at the end ($\tau \sim 4 \times 10^{-9}$ s)---at this stage, the mobilities are high and have the same forms both in the volume and at the boundary:

\begin{equation*}
\overline{V} = \frac{\alpha}{\delta R} D_{b}^{*}
\end{equation*}



for both (66) and (62).

II. The stage between curves ($\alpha$) and ($\beta$), Fig. 4---at this stage, which lasts $ \sim 2.6 \times 10^{-8}$ s, according to the form of $g(C_{a}^{l})$, the processes occurring at the boundary are sufficiently suppressed (Fig. 4), but the mobilities are the same in the volume as those at the first stage up to $\beta$. Thus, the values of $\overline{V}$ in (66) are much more than those in (62).



III. The stage after reaching the curve ($\beta$)---the processes are suppressed both in the volume and in the near-boundary zone. The duration of the third stage does not matter for the estimation of $Z$ because if the mobilities are equal, their dependence on the duration disappears. The calculation of $Z$ in this case yields values of $Z \geq 1.4$.

It implies that it is possible to select such functional dependences of activity, solubility, mobility parameters, and solute element concentrations which change the solidification mechanism and stop the formation of clusters.
 
\section{Conclusion}





We would like to draw attention to the fact that, strictly speaking, solute-enriched clusters are radiation-induced only at the stage of thermal spike formation.

All further development is a local first order phase transition involving solute drag, which renders this process fundamentally different from the usual radiation processes such as radiation growth, swelling, radiation-induced creep, and radiation-induced segregation.

The analysis conducted enables complete formularization of a self-consistent problem on the determination of the concentration of the solute element ''$a$'' consisting of

- the non-linear diffusion equation (56);

- an initial condition;

- a boundary condition in the form of a flux through the boundary, specified by expressions (51), (53), and (55) and depending on the range of Ca values in the near-boundary zone; and

 - the boundary motion rates (50), (52), (54).
 


This, in turn, enables us to describe the scenario of liquid phase development and select the parameters $a(C_{a}^{l}), \frac{D_{a}^{*}}{D_{b}^{*}}, and \overline{C}_{a}, E_{a}^{sol,s}$ (for actually choosing the solute and its concentration) which make it possible to adopt a mode that excludes solute-enriched cluster formation.

It has been demonstrated that by selecting the solute properties, it is possible to change the thermal spike molten zone solidification mechanism, and thus, suppress the grain body hardening mechanism which causes the formation of solute-enriched clusters.

\section*{Acknowledgement}
The authors acknowledge Dr. O. Telkovskaya from NRC "Kurchatov Institute" and Dr. S. Panyukov from P. N. Lebedev Physical Institute for their fruitful discussions and assistance with this work.

The work was supported by the NRC "Kurchatov Institute" ($N$ 1579).

This work has been carried out by using the computing resources of the federal collective usage centre Complex for Simulation and Data Processing for Mega-Science Facilities at NRC Kurchatov Institute, \url{http://ckp.nrcki.ru}.

\newpage

\section*{Appendix 1}
\appendix



 
Numerical method of solution of equations:

Formulas (56), (57), and (58) yield a consistent system of two equations: one of them is a partial derivative evolution equation, and the second is an ordinary differential equation. 

According to the formula $J_{a} = -V C_{a} + I_{a}$, it is also necessary to calculate the leakage of "$a$" component to the solid phase $I_{a}$ for complete closure of the system. 

Formulas (51), (53), and (55) provide the dependence of $I_{a}$ on the cuprum concentration at the liquid phase boundary. 




Since the cuprum concentration never reached the application range in the calculations based on (55), it was only the relations (51) and (53) which were practically used. At the value $C_{a}^{A}$, the transition from (51) to (53) yielded a functional step while computing in the small neighbourhood of $C_{a}^{A}$, where linear interpolation was used.

In all the calculations, the values of $I_{a}$ were small in comparison with $C_{a}$, therefore, these approximations did not play a significant role. 

Since the system exhibits spherical symmetry, it is possible to simplify the problem (56)-(58):

\begin{equation*}
\frac{\partial C_{a}}{\partial t} = \frac{1}{r^{2}} \frac{\partial}{\partial r} \left(r^{2} D_{a}(C_{a}) \frac{\partial C_{a}}{\partial r} \right)
\end{equation*}

\begin{equation*}
J_{a}(R,t) = -D_{a} \frac{\partial C_{a}}{\partial r}
\end{equation*}

\begin{equation*}
\frac{d R}{d t} = -V(C_{a}(R,t))
\end{equation*}



It is convenient to introduce a moving coordinate system in the calculations for spatial one-dimensional problems with moving boundaries \cite{1995CW}, \cite{Sam}. 

In the present study, the following coordinates were used.

\begin{equation}
p = \frac{r}{R} ;  \tau=t														
\end{equation}


whence (by using (57))

\begin{equation}
\frac{\partial}{\partial r} = \frac{1}{R} \frac{\partial}{\partial p} ; \frac{\partial}{\partial t} = \frac{\partial}{\partial \tau} + \frac{p V}{R} \frac{\partial}{\partial p}
\end{equation}


The system takes the following form when represented in these coordinates ($p$ is always in [0,1]):

\begin{equation}
\frac{\partial C_{a}}{\partial \tau} = \frac{1}{R^{2} p^{2}} \frac{\partial}{\partial p} \left(D_{a} p^{2} \frac{\partial C_{a}}{\partial p} \right) - \frac{pV}{R} \frac{\partial C_{a}}{\partial p} 
\end{equation}

\begin{equation}
- J_{a}(1,\tau) = D_{a} (C_{a}(1, \tau)) \frac{\partial C_{a}}{R \partial p} (1, \tau) = F(C_{a}(1, \tau))
\end{equation}

\begin{equation}
\frac{d R}{d \tau} = -V (C_{a}(1, \tau))
\end{equation}


The form of (A3) with open brackets is also used for the point $p = 1$:

\begin{equation}
\frac{\partial C_{a}}{\partial \tau} = \frac{D_{a}}{R^{2}} \frac{\partial^{2} C_{a}}{(\partial p)^{2}} + \frac{1}{R^{2}} \frac{\partial D_{a}}{\partial p} \frac{\partial C_{a}}{\partial p} + \left( \frac{ 2D_{a}}{R^{2} p} - \frac{p V}{R} \right)  \frac{\partial C_{a}}{\partial p} 
\end{equation}




A uniform grid $h = 1/M$, $p_{i} = hi = i/M$, $(i=0,1,...,M)$ was used for digitalization involving spatial variables.

We used difference approximation of the diffusion term from the paper \cite{1995CW}, which provides recommendations for one-dimensional parabolic non-linear equations.

A zero flow is supposed for the sphere centre, therefore, it is considered that $C_{a}(0,\tau) \equiv C_{a}(h,\tau)$, and the function is actually considered for the grid with indices $1,2..., M.$




Further, $C_{a}(p_{i},\tau)$ will be designated as $C_{ai}$ and $D(C_{ai})$ as $D_{i}$. Besides, the equations $p_{i}=i/M$ are used.

Let us introduce the designation $Q_{i}=i^{2} D_{i}$.

After digitalization by the spatial variable, we obtain a system of ordinary differential equations of the form

\begin{equation}
 \frac{d C_{ai}}{d \tau} = f (C_{a1}, ... ,C_{aM}, R, \tau), 1 \leq i \leq M \atop\frac{d R}{d \tau} = - V
\end{equation}


The scheme in which full by sphere flow-over between the nodes $i-1$ and $i$ is approximated by the expression

\begin{equation*}
-\frac{p_{i-1}^{2} D_{i-1} + p_{i}^{2} D_{i}}{2} \times \frac{C_{ai} - C_{ai-1}}{h}
\end{equation*}


which describes the flow-over balance, except at the extreme points. Therefore, it is used for the approximation of the diffusion term. 

\begin{equation*}
\frac{1}{R^{2} p^{2}} \frac{\partial}{\partial p} \left( D_{a} p^{2} \frac{\partial C_{a}}{\partial p} \right)
\end{equation*}


for all the nodes except the $M$th. We use the equality $C_{a0} = C_{a1}$ for the first node. For the approximation term

\begin{equation*}
-\frac{p V}{R} \frac{\partial C_{a}}{\partial p}
\end{equation*}



we use the central difference derivative for the points $1,2,...M-1$.

Considering that

\begin{equation*}
\frac{p_{i}^{2} D_{i}}{p_{j}^{2}} = \frac{i^{2} D_{i}}{j^{2}} = \frac{Q_{i}}{j^{2}}
\end{equation*}


we obtain the following expressions for the right-hand side parts of the ODE system $f_{i}$:

\begin{equation}
\frac{\partial C_{a1}}{\partial \tau} = f_{1} = \frac{Q_{2} + Q_{1}}{2 R^{2} h^{2}} 
\left(C_{a2}-C_{a1} \right) - \frac{V}{R} \left(C_{a2}-C_{a1} \right)
\end{equation}

\begin{equation}
\frac{\partial C_{ai}}{\partial \tau} = f_{i} = \frac{(Q_{i-1} + Q_{i})(C_{ai-1}-C_{ai}) + (Q_{i+1} + Q_{i})(C_{ai+1}-C_{ai})}{2 R^{2} h^{2} i^{2}} -
\end{equation}

\begin{equation*}
- \frac{i V}{2 R} (C_{ai+1}- C_{ai-1}) , (i < 1 \leq M)
\end{equation*}


For $i = M$, i.e.  $p_{i} = 1$, the form (A6) is used; in this case, it takes the form

\begin{equation*}
\frac{\partial C_{a}}{\partial \tau} = \frac{D_{a}}{R^{2}} \frac{\partial^{2} C_{a}}{(\partial p)^{2}} + \frac{1}{R^{2}} \frac{\partial D_{a}}{\partial p} \frac{\partial C_{a}}{\partial p} + \left( \frac{ 2D_{a}}{R^{2} p} - \frac{p V}{R} \right)  \frac{\partial C_{a}}{\partial p} 
\end{equation*}


Accordingly, it is possible to write $f_{M}$ as a sum of 3 addends

\begin{equation}
-\frac{d C_{aM}}{d \tau} = f_{M} = Z_{1} + Z_{2} + Z_{3}
\end{equation}


It follows from Taylor's expansion that

\begin{equation*}
\frac{\partial^{2} C_{a}}{(\partial p)^{2}}(1) = \frac{2}{h^{2}} (C_{aM-1}-C_{aM}) + \frac{2}{h}\frac{\partial C_{a}}{\partial p}(1) + O(h)
\end{equation*}


Formula (A4) yields

\begin{equation*}
\frac{\partial C_{a}}{\partial p}(1) = \frac{F(C_{M})}{D_{M}} R
\end{equation*}


From this, we obtain the approximation

\begin{equation}
Z_{1} = \frac{2 D_{M}}{R^{2} h^{2}}(C_{aM-1}-C_{aM}) + \frac{2}{R h} F(C_{aM})
\end{equation}


As the point where $M+1$ is absent, we use the left difference derivative for the first derivative in the rest of the addends:

\begin{equation}
Z_{2} = \frac{D_{M} - D_{M-1}}{R^{2} h^{2}}(C_{aM}-C_{aM-1})
\end{equation}

\begin{equation}
Z_{3} = \frac{2 D_{M} - V R}{R^{2} h} (C_{aM}-C_{aM-1})
\end{equation}



It is clear that the formulas are reduced to a more convenient computational form in the realization of the program.

The abovementioned approximations are used to calculate the functions $F(C_{a})$ and $V(C_{a})$.



It is worth noting that it follows immediately from the given formulas that if we formally designate $y_{i} =C_{ai}$, $i \leq M$, $y_{M=1} = R$ and $f_{M+1}=-V$, the Jacobi matrix $\left(\frac{\partial f_{i}}{\partial y_{j}} \right)$ has only one diagonal adjoining the main diagonal in the lower triangle. In other words, solving a system of linear equations by using such a matrix does not require too much calculations.

We also observe that the formulas (51), (53), and (57) are such that calculations of the partial derivatives involving various $y_{j}$ only require the presence of a function for calculating the value
$\frac{\partial D_{a}}{\partial C_{a}}(C_{a}) = \frac{d D_{a}}{d C_{a}}$, because $D_{a}$ depends only on $C_{a}$.



At first, several variants of the dependence $D_{a}(C_{a})$ were used in the calculations. The abovementioned property enabled us to change only two functions $\frac{d D_{a}}{d C_{a}}$ and $\frac{d D_{a}}{d C_{a}}$ for calculating $D_{a}(C_{a})$.




Since the present program uses implicit schemes, it has an option in terms of when the user can input a subprogram for the calculation of the Jacobi matrix, which accelerates the computation process. It is the availability of this option that facilitated the use of not only the function $D_{a}(C_{a})$ but also $\frac{d D_{a}}{d C_{a}}$; furthermore, the obtained matrix was only a Hessenberg matrix.

During the calculation, we conducted tests to control the accuracy. In the reduced difference approximations at points $i < M$, as well as at $ i = M+1$ (equation for $R$), we provide an accuracy of up to $O(h^{2})$, whereas, at the point $M$, it is only $O(h)$. 

 


It is possible to improve the scheme with respect to the terms $Z_{2}$ and $Z_3$, but the order of accuracy remains the same. This disadvantage is slightly mitigated by the smallness of the coefficients of the $h$ order terms. 

Control calculations with different numbers of intervals $M$ were conducted. They showed that an increase in $M$ above 400 (the main variant) does not have a considerable influence.

We also performed calculations with an improved difference scheme for the extreme point $M$, but that did not yield considerable refinement either.

\section*{Appendix 2}

$E_{PKA}$ - energy of a primary knock-on atom;

$D_{a}^{*}, D_{b}^{*}$ - tracer diffusion coefficient for the components ''$a$'' and ''$b$'', respectively;

$V(C_{a},t)$  - sol - liq boundary motion rate;

$a_{a}, a_{b}$ - activities;

$\mu_{a}^{l}, \mu_{b}^{l}, \mu_{a}^{s}, \mu_{b}^{s}$ - chemical potentials of the components ''$a$'' and ''$b$'' in the liquid and solid phases;

$I_{a}, I_{b}$ - fluxes of the components ''$a$'' and ''$b$'' from the liquid phase to the solid phase based on the approximation of a ''thin boundary'' (see (6), (8)); refined (see (51), (53), (55));

$J_{a}, J_{b}$ - fluxes of ''$a$'' and ''$b$'' from the liquid phase to the boundary zone;

$j_{a}, j_{b}$ - fluxes with regard to the boundary zone (minus the drift, see (12));

$\overline{V}$ - the sol-liq boundary motion rate, considering the drift in the near-boundary region $\raisebox{.5pt}{\textcircled{\raisebox{-.9pt} {B}}}$ (20);

$\kappa \equiv \frac{C_{a}^{s}}{C_{a}^{l}}$ -  solute drag measure (34), (45);

$\delta_{eq} \equiv \frac{a_{a}^{s}}{a_{a,0}^{s}}$  - equilibrium deviation measure (36), (45);

$a_{a,0}^{s}$ - $\Delta \mu_{a}^{s-l} = 0$ from conditions (36), (45).

\section*{Appendix 3}

$\delta E_{a} \equiv E_{sol,a}^{s} - \widetilde{h}_{a} \left(\frac{T_{m} - T}{T_{m}} \right) \approx 0.3 eV$

$\widetilde{\sigma} = 7.5 \cdot 10^{-2} eV$, - sol - liq boundary energy per atom;

$\widetilde{h}_{a,b} = 1.5 \cdot 10^{-1} eV$, melting enthalpy per atom of iron;

$T_{m} = 1810 K$, - melting temperature of iron;

$T = 580 K$, - process temperature;

$kT = 0.05 eV$;

$\theta = \frac{T_{m} - T}{T_{m}} = 0.68$;

$\alpha = f \frac{6 a_{0} \delta R}{l^{2}} exp \left\lbrace \frac{\delta E_{b} - \widetilde{s} T}{kT} \right\rbrace = 4.05$;

$a_{0} \approx 2.6 A$  - lattice parameter;

$\delta R \approx (3 \div 4) a_{0}$; - interface ''thickness'';

$l\approx 2 a_{0}$;  - atom jump length across the interface;

$f\approx 0.25$; - surface roughness measure;

$\delta E_{b} = 0.1 eV$;

$\widetilde{s} T = 0.05 eV$;

$R_{max} = 5 \cdot 10^{-7}$ cm; - initial radius of the molten zone;

$s_{0} \equiv 3 \left( \frac{4 \pi}{3} \right)^{1/3} \approx 4.8$;

$\Omega ^{Fe} = 1.17 \cdot 10^{-23}$ cm$^{3}$; - atomic volume of iron;

$D_{b}^{*} = 10^{-6}$  cm$^{2}$/s; - self-diffusion coefficient of iron;

$\gamma_{max} \approx 5.3 \cdot 10^{-19}$ cm$^{3}$; - initial volume of the molten zone;





\newpage
\section{Figures, tables, references}
\label{sec:others}



\subsection{Figures}



\begin{figure}[h]
\center{\includegraphics[width=1\linewidth]{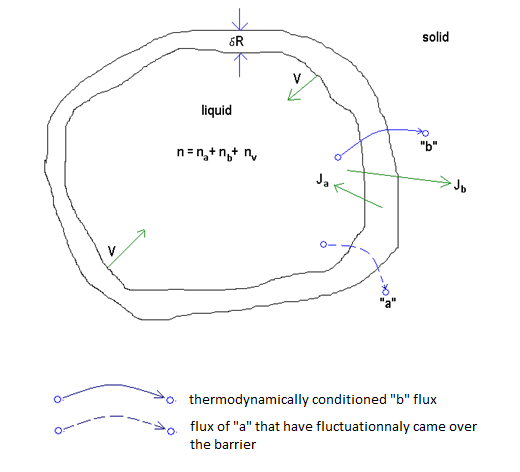}}
\caption{Liq-sol transition for $T < T_{m}$ by ''jump'' mechanism;
$\delta R$ is the sol-liq boundary thickness}
\label{ris:fig1}
\end{figure}

\newpage
\begin{figure}[h]
\center{\includegraphics[width=1\linewidth]{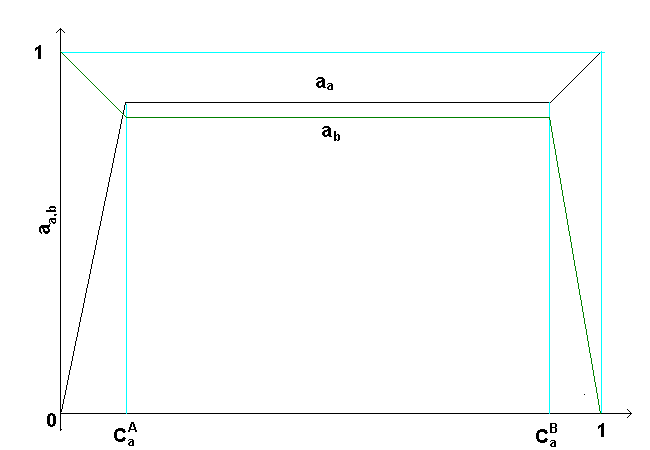}}
\caption{Dependences of the activities of "$a$" and "$b$" components in different concentration ranges}
\label{ris:fig3}
\end{figure}

\newpage
\begin{figure}[h]
\center{\includegraphics[width=1\linewidth]{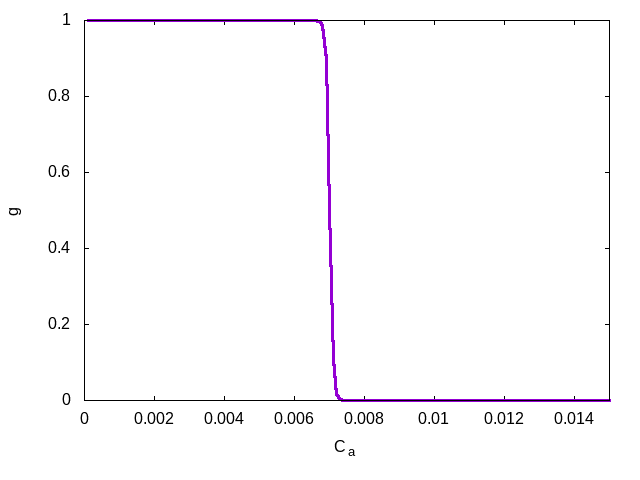}}
\caption{Graph of function g}
\label{ris:Fig4}
\end{figure}

\newpage
\begin{figure}[h]
\center{\includegraphics[width=1\linewidth]{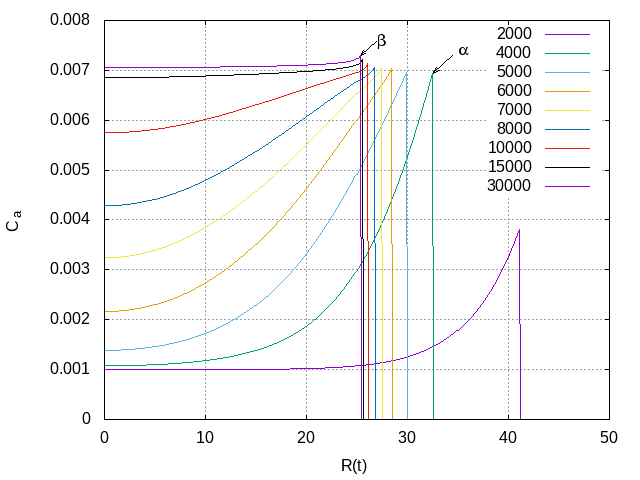}}
\caption{Dependences of the concentration profiles on time}
\label{ris:figout}
\end{figure}

\newpage
\begin{figure}[h]
\center{\includegraphics[width=1\linewidth]{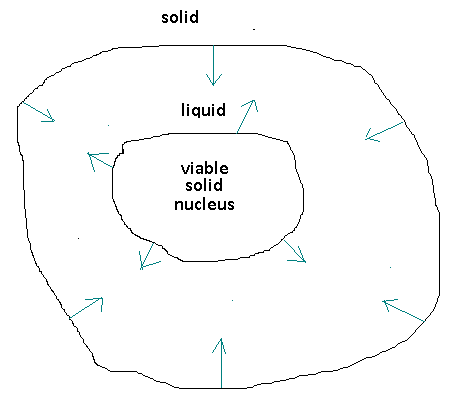}}
\caption{Change in the phase transformation mechanism}
\label{ris:fig5}
\end{figure}

\newpage
\subsection{Tables}

\begin{table}[h]
\caption{\label{tab:tablica2} Estimation of the degree of solute drag}
\begin{center}
\begin{tabular}{|c|c|}

\hline
$D_{a}=2D_{b}$ & $\kappa=1.5 \cdot 10^{-3}$ \\
\hline
$D_{a}=3D_{b}$ & $\kappa=7.4 \cdot 10^{-4}$ \\
\hline
$D_{a}=4D_{b}$ & $\kappa=5.5 \cdot 10^{-4}$ \\
\hline
\end{tabular}
\end{center}
\end{table} 

\begin{table}[h]
\caption{\label{tab:tablica3} Estimation of the equilibrium deviation measure}
\begin{center}
\begin{tabular}{|c|c|}

\hline
$D_{a}=2D_{b}$ & $\delta_{eq} \cong 7.4$ \\
\hline
$D_{a}=3D_{b}$ & $\delta_{eq} \cong 3.8$ \\
\hline
$D_{a}=4D_{b}$ & $\delta_{eq} \cong 2.7$ \\
\hline
\end{tabular}
\end{center}
\end{table} 

\newpage
\begin{table}[h]
\caption{\label{tab:tablica4} Estimations of $C_{a}^{**l}$}
\begin{center}
\begin{tabular}{|c|c|}

\hline
$\frac{D_{a}^{*}}{D_{b}^{*}}=2$ & $C_{a}^{**l} \geq 1 \geq C_{a}^{l}$ \\
\hline
$\frac{D_{a}^{*}}{D_{b}^{*}}=2.4$ & $C_{a}^{**l}=C_{a}^{l}=0.86$ \\
\hline
$\frac{D_{a}^{*}}{D_{b}^{*}}=3$ & $C_{a}^{**l}=0.75$ \\
\hline
$\frac{D_{a}^{*}}{D_{b}^{*}}=4$ & $C_{a}^{**l}=0.67$ \\
\hline
\end{tabular}
\end{center}
\end{table} 

\begin{table}[h]
\caption{\label{tab:tablica5} Estimations of $C_{a}^{s}$}
\begin{center}
\begin{tabular}{|c|c|}

\hline
$\frac{D_{a}}{D_{b}}=3$ & $C_{a}^{s} \leq \frac{1}{2} C_{a}^{l}$ \\
\hline
$\frac{D_{a}}{D_{b}}=4$ & $C_{a}^{s} \leq \frac{2}{3} C_{a}^{l}$ \\
\hline
\end{tabular}
\end{center}
\end{table}

\newpage



\bibliographystyle{unsrt}  


\end{document}